\documentclass[aps,prd,eqsecnum,preprint,tightenlines,nofootinbib]{revtex4-1}
\usepackage{graphicx,latexsym}

\newcommand{\bea}{\begin{eqnarray}}
\newcommand{\eea}{\end{eqnarray}}
\newcommand{\be}{\begin{equation}}
\newcommand{\ee}{\end{equation}}

\newcommand{\Pminus}{{\cal P}^-}

\newcommand{\Nmax}{N_{\rm max}}

\begin{document}

\title{Light-front versus equal-time quantization in $\phi^4$ theory%
\footnote{Based on a talk contributed to the
Lightcone 2016 workshop, Lisbon, Portugal, 
September 5-8, 2016.}
}

\author{S.S. Chabysheva}

\affiliation{Department of Physics and Astronomy\\
University of Minnesota-Duluth \\
Duluth, Minnesota 55812}

\date{\today}

\begin{abstract}
There is a discrepancy between light-front and equal-time values
for the critical coupling of two-dimensional $\phi^4$ theory.
A proposed resolution is to take into account the difference
between mass renormalizations in the two quantizations.
This distinction was first discussed by M. Burkardt.  It
prevents direct comparison of bare parameters; however, a
method proposed here allows calculation of the difference
and thereby resolves the discrepancy.  We also consider
the consequences of allowing a sector-dependent constituent mass.
\end{abstract}

\maketitle

\section{Introduction} \label{sec:intro}

Previous calculations~\cite{phi4sympolys,Frascati} have 
shown that there is a systematic difference between light-front
and equal-time values for the critical coupling of
$\phi^4$ theory.  Here we discuss the resolution of
this by considering the difference in mass renormalizations
as originally suggested by Burkardt~\cite{SineGordon}.
However, we also show that the expected behavior of the
probability for higher Fock sectors, that they
should grow dramatically as the critical coupling
is approached, is not observed. This interferes with
the estimation of the mass renormalization.  As a means
to resolve this, we consider use of sector-dependent 
constituent mass in the solution of the Hamiltonian
eigenvalue problem.  Many of the details of the methods used,
particularly for this eigenvalue problem, are presented
by Hiller~\cite{HillerLC16}.

\begin{figure}[ht]
\centering
\includegraphics[width=12cm]{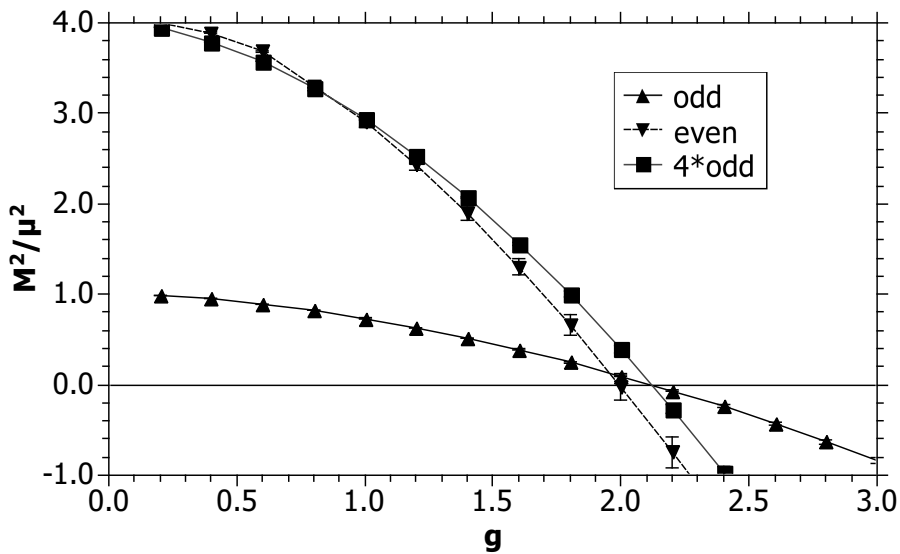}
\caption{\label{fig:critcoup} 
The lowest masses for the odd and even eigenstates
of $\phi^4$ theory, from~\protect\cite{phi4sympolys}.  The 
threshold for two-particle states, at four times the 
mass-squared of the odd case, is also shown.
}
\end{figure}

Our light-front estimate of the critical coupling is
based on calculations of the lowest massive eigenstates
as a function of the coupling for both the odd and even
sectors.  The values of the eigenmass squared are 
plotted in Fig.~\ref{fig:critcoup}.
The intersections of the spectra with zero indicate
a critical coupling of $g=2.1\pm0.05$.
This value is compared with values from other calculations
in Table~\ref{tab:critcoup}.
\begin{table}[ht]
\caption{\label{tab:critcoup}
Critical coupling values, adapted 
from~\protect\cite{RychkovVitale}, with use of a slightly
different definition of the dimensionless
coupling $\bar{g}=\frac{\pi}{6}g$.
The first two values were computed
in light-front quantization and the remainder in equal-time
quantization.}
\begin{center}
\begin{tabular}{lll}
\hline \hline
Method &  $\bar{g}_c$ & Reported by \\
\hline
Light-front symmetric polynomials & $1.1\pm0.03$ & this work \\
DLCQ  & 1.38 & Harindranath \& Vary~\protect\cite{VaryHari} \\
\hline
Quasi-sparse eigenvector & 2.5 & 
     Lee \& Salwen~\protect\cite{LeeSalwen} \\
Density matrix renormalization group & 2.4954(4) & 
     Sugihara~\protect\cite{Sugihara} \\
Lattice Monte Carlo & 2.70$\left\{\begin{array}{l} +0.025 \\ -0.013\end{array}\right.$ & 
     Schaich \& Loinaz~\protect\cite{SchaichLoinaz} \\
                    & $2.79\pm0.02$ & Bosetti {\em et al.}~\protect\cite{Bosetti} \\
Uniform matrix product & 2.766(5) & 
     Milsted {\em et al.}~\protect\cite{Milsted} \\
Renormalized Hamiltonian truncation & 2.97(14) & 
     Rychkov \& Vitale~\protect\cite{RychkovVitale} \\
\hline \hline
\end{tabular}
\end{center}
\end{table}

There is clearly a systematic difference between results
for light-front quantization and equal-time quantization.
In the following section we describe the resolution of
this difference in terms of a shift in the effective
constituent mass.  We also discuss, in Sec.~\ref{sec:secdep},
the implementation of a sector-dependent constituent mass
as an attempt to improve the calculation of the shift.
A summary is given in Sec.~\ref{sec:summary}.

\section{Mass renormalization} \label{sec:mass}

The bare mass is renormalized by tadpole contributions in 
equal-time (ET) quantization but not in light-front (LF)
quantization.  With the Lagrangian written as 
${\cal L}=\frac12(\partial_\mu\phi)^2-\frac12\mu_0^2\phi^2-\frac{\lambda}{4!}\phi^4$,
this results in a difference that can be expressed as~\cite{SineGordon}
\be
\mu_{\rm LF}^2=\mu_{\rm ET}^2
   +\lambda\left[\langle 0|\frac{\phi^2}{2}|0\rangle
       -\langle 0|\frac{\phi^2}{2}|0\rangle_{\rm free}\right].
\ee
where the vacuum expectation values (VEV) of $\phi^2$ resum the tadpole
contributions.  Here the subscript {\em free} indicates the VEV with
zero coupling.

To calculate the VEV's, we regulate them with a point splitting
by $(\epsilon^+,\epsilon^-)$ and introduce a decomposition of the identity
in terms of the eigenstates $|\psi_n(P)\rangle$ of the light-front Hamiltonian $\Pminus$:
\be
\langle 0|\frac{\phi^2}{2}|0\rangle\rightarrow 
   \frac12\langle 0|\phi(\epsilon^+,\epsilon^-)\int_0^\infty dP\sum_n|\psi_n(P)\rangle
      \langle\psi_n(P)|\phi(0,0)|0\rangle.
\ee
The operator $\phi(\epsilon^+,\epsilon^-)$ is obtained by evolving forward in
light-front time $x^+\equiv t+z$ from zero to $\epsilon^+$, so that
$\phi(\epsilon^+,\epsilon^-)=e^{i\Pminus\epsilon^+/2}\phi(0,\epsilon^-)e^{-i\Pminus\epsilon^+/2}$.

The necessary matrix elements for the $n$th bound state are
\be
\langle\psi_n(P)|\phi(0,0)|0\rangle
  =\langle 0|\psi_{n1}^*a(P)\int\frac{dp}{\sqrt{4\pi p}}a^\dagger(p)|0\rangle
  =\frac{\psi_{n1}^*}{\sqrt{4\pi P}}
\ee
and 
\be
\langle 0|\phi(\epsilon^+,\epsilon^-)|\psi_n(P)\rangle=
\langle 0|\int\frac{dp}{\sqrt{4\pi p}}a(p)
  e^{-ip\epsilon^-/2}e^{-iM_n^2\epsilon^+/2P}\psi_{n1}a^\dagger(P)|0\rangle
 =\frac{\psi_{n1}}{\sqrt{4\pi P}}e^{-i(P\epsilon^- +M_n^2\epsilon^+/P)/2}.
\ee
Similarly, for the one-particle free state, the matrix elements are
\be
\langle0|a(P)\phi(0,0)|0\rangle
  =\langle 0|a(P)\int\frac{dp}{\sqrt{4\pi p}}a^\dagger(p)|0\rangle
  =\frac{1}{\sqrt{4\pi P}} 
\ee
and
\be
\langle 0|\phi(\epsilon^+,\epsilon^-)a^\dagger(P)|0\rangle=
   \langle 0|\int\frac{dp}{\sqrt{4\pi p}}a(p)
  e^{-ip\epsilon^-/2}e^{-i\mu^2\epsilon^+/2P}a^\dagger(P)|0\rangle
=\frac{1}{\sqrt{4\pi P}}e^{-i(P\epsilon^- +\mu^2\epsilon^+/P)/2}.
\ee
On substitution of these matrix elements, the VEV's can be 
written as
\be
\langle 0|\frac{\phi^2}{2}|0\rangle=\frac12\sum_n\int_0^\infty dP \frac{|\psi_{n1}|^2}{4\pi P}
       e^{-i(P\epsilon^- +M_n^2\epsilon^+/P)/2}  \;\;
\mbox{and} \;\;
\langle 0|\frac{\phi^2}{2}|0\rangle_{\rm free}=\frac12\int_0^\infty dP \frac{1}{4\pi P}
       e^{-i(P\epsilon^- +\mu^2\epsilon^+/P)/2}.
\ee
With use of $1=\sum_n|\psi_{n1}|^2$, the difference can be reduced to
\be
\langle 0|\frac{\phi^2}{2}|0\rangle-\langle 0|\frac{\phi^2}{2}|0\rangle_{\rm free}
=\sum_n \frac{|\psi_{n1}|^2}{8\pi}\int_0^\infty \frac{dP}{P} e^{-iP\epsilon^-/2}
\left[e^{-i\frac{M_n^2\epsilon^+}{2P}}-e^{-i\frac{\mu^2\epsilon^+}{2P}}\right]
\ee

The mass shift is then proportional to
\be
\langle 0|\frac{\phi^2}{2}|0\rangle-\langle 0|\frac{\phi^2}{2}|0\rangle_{\rm free}
=\sum_n \frac{|\psi_{n1}|^2}{4\pi}\left[K_0(M_n\sqrt{-\epsilon^2+i\eta})
                 -K_0(\mu\sqrt{-\epsilon^2+i\eta})\right],
\ee
where $\eta$ is a convergence factor.  From the asymptotic behavior of
the Bessel function $K_0(z)\rightarrow-\ln(z/2)-\gamma$, we obtain
\be \label{eq:Delta}
\langle 0|\frac{\phi^2}{2}|0\rangle-\langle 0|\frac{\phi^2}{2}|0\rangle_{\rm free}
=-\sum_n \frac{|\psi_{n1}|^2}{4\pi}\ln\frac{M_n}{\mu_{\rm LF}}\equiv -\Delta/4\pi,
\ee
Finally, the relationship between the masses is
\be 
\mu_{\rm LF}^2=\mu_{\rm ET}^2-\frac{\lambda}{4\pi}\Delta \;\;
\mbox{or} \;\;
\frac{\mu_{\rm ET}^2}{\mu_{\rm LF}^2}=1+g_{\rm LF}\Delta.
\ee
This implies coupling constant and mass ratios of
\be \label{eq:couplingratio}
g_{\rm ET}=\frac{g_{\rm LF}}{\mu_{\rm ET}^2/\mu_{\rm LF}^2}=\frac{g_{\rm LF}}{(1+g_{\rm LF}\Delta)}\;\;
\mbox{and} \;\;
\frac{M^2}{\mu_{\rm ET}^2}
      =\frac{1}{1+g_{\rm LF}\Delta}\frac{M^2}{\mu_{\rm LF}^2}
\ee

\begin{figure}[ht]
\vspace{0.2in}
\centerline{\includegraphics[width=12cm]{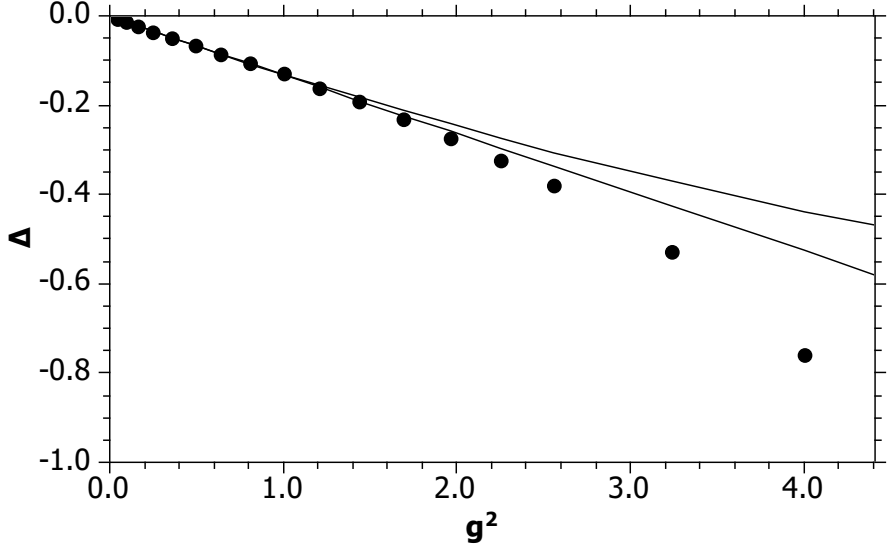}}
\caption{\label{fig:shift}
The renormalization shift $\Delta$ as a function of the square of the 
dimensionless coupling $g$, as shown in \protect\cite{phi4sympolys}.
The points displayed are obtained as extrapolations in the polynomial
basis size.  The lines are linear and quadratic fits to shifts below 
$g=1$, extrapolated to the region of the critical coupling.}
\end{figure}

The shift $\Delta$ is plotted in Fig.~\ref{fig:shift}.  As the
critical coupling is approached, the shift does not behave as
it should.  Fits are then done for coupling values much less
than the critical value, in order to extrapolate.
The extrapolations of the shift then yield
$\Delta(g=2.1)=-0.47\pm0.12$.  The latest equal-time value
for the critical coupling~\cite{RychkovVitale},
$g_{\rm ETc}=\frac{6}{\pi}2.97=5.67$, implies a shift
of $(g_{\rm LFc}/g_{\rm ETc}-1)/g_{\rm LFc}=-0.30$,
which is consistent.

\begin{figure}[ht]
\vspace{0.2in}
\centerline{\includegraphics[width=12cm]{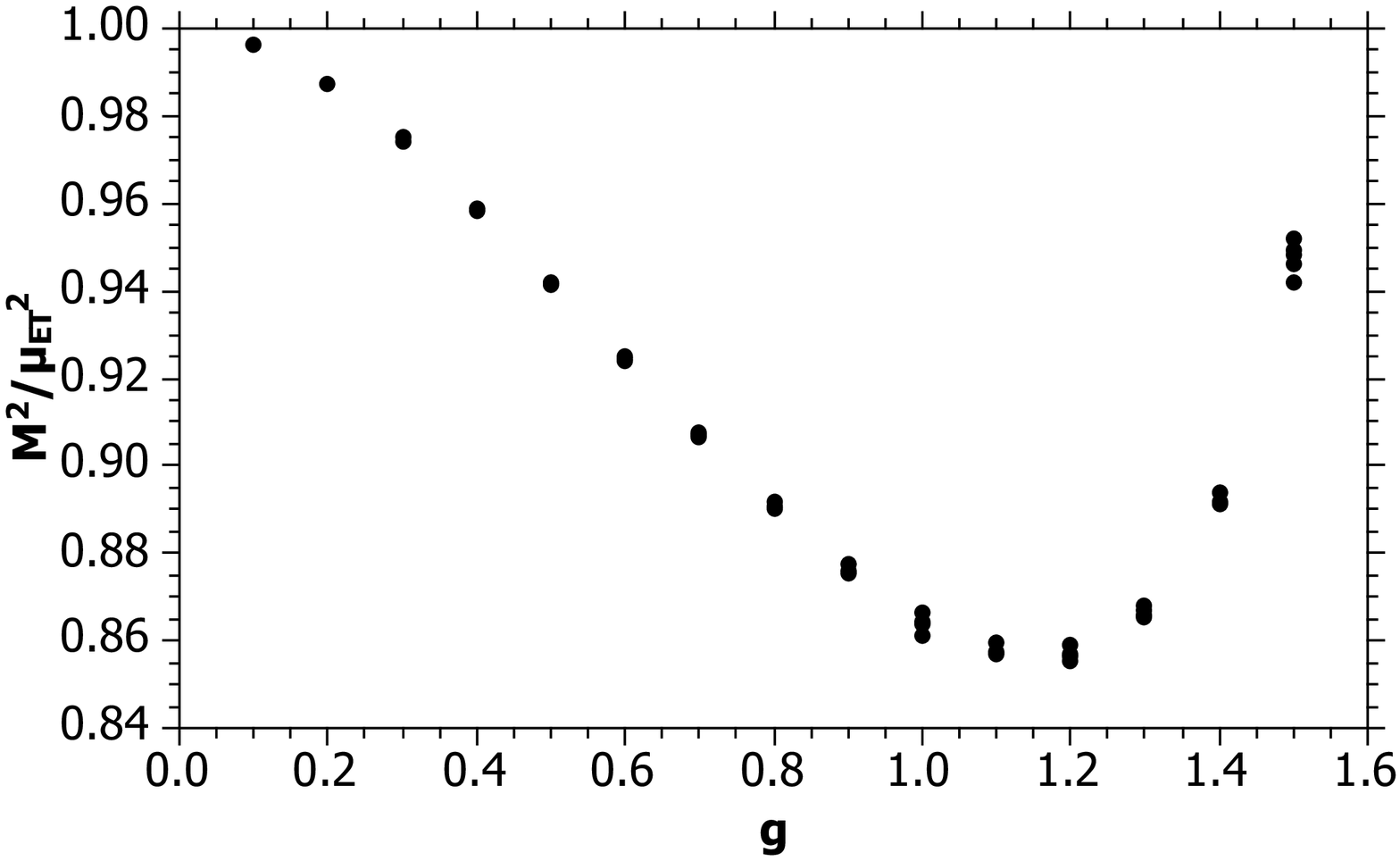}}
\caption{\label{fig:ETmass}
Lowest equal-time mass eigenvalues for odd numbers of constituents
plotted versus the dimensionless light-front coupling $g$,
from \protect\cite{phi4sympolys}.
Different points at the same $g$ value correspond to
different truncations of the polynomial basis size.
}
\end{figure}

The criterion for the set of coupling values used is determined
by examining the behavior of the predicted equal-time values
for the mass squared, based on the relationship in 
(\ref{eq:couplingratio}), as shown in Fig.~\ref{fig:ETmass}.
Above $g=1$ the equal-time masses begin to increase rather
than continue the proper decrease.

The origin of the increase in the mass, and the improper 
behavior of the shift near the critical coupling, is in the
finiteness of the relative probabilities for higher Fock states,
displayed in Fig.~\ref{fig:relprobsecdep}.
As the critical coupling is approached there is only a continued
gradual increase, implying that the one-body probability $|\psi_{11}|^2$
remains nonzero.  This causes the shift $\Delta$ to diverge
as the eigenmass $M_1$ goes to zero.

\section{Sector-dependent constituent mass} \label{sec:secdep}

With a sector-independent constituent mass $\mu$, the invariant 
mass of a higher Fock state is quite large and such states
are then naturally suppressed in any calculation with a 
Fock-space truncation.  An obvious way to avoid this suppression
is to use a sector-dependent 
mass~\cite{SecDep1,SecDep2,SecDep3,SecDep4,SecDep5,SecDep6} $\mu_m$.  
The matrix eigenvalue problem given in \cite{HillerLC16} can then be written as
\bea 
\sum_{n'i'}\left[\tilde{\mu}_m^2 T^{(m)\prime}_{ni,n'i'}\right.
   &+& \left.V^{(m,m)\prime}_{ni,n'i'}\right]c^{(m)\prime}_{n'i'}
   + \sum_{n'i'} V^{(m,m+2)\prime}_{ni,n'i'} c^{(m+2)\prime}_{n'i'}  \\
 &+& \sum_{n'i'} V^{(m,m-2)\prime}_{ni,n'i'} c^{(m-2)\prime}_{n'i'}
   =\tilde{M}^2 c_{ni}^{(m)\prime},
\eea
with $\tilde{\mu}_m\equiv\mu_m\sqrt{4\pi/\lambda}$ and $\tilde{M}\equiv M\sqrt{4\pi/\lambda}$.
The sector-dependent mass then allows for the fact that a Fock-space
truncation forces self-energy corrections to be different in each
Fock sector; in particular, in the highest sector there is no self-energy
correction.
When the Fock-space truncation is removed, in the $\Nmax\rightarrow\infty$ limit,
the dimensionless sector-dependent mass $\tilde\mu_m$ becomes
$\tilde{\mu}\equiv\pm 4\pi\mu^2/\lambda$.  This convergence to the 
sector-independent mass happens first in the lowest sector; therefore,
the dimensionless coupling $g\equiv\frac{\lambda}{4\pi\mu^2}$ can be 
extracted as $g\simeq 1/|\tilde\mu_1^2|$ and the dimensionless
eigenmass as $M^2/\mu^2=g\tilde M^2=\tilde M^2/|\tilde\mu_1^2|$.

The sector-dependent case is no longer an explicit eigenvalue problem.
The sector-dependent masses $\tilde\mu_m$ must be computed recursively for
a given value of $\tilde M$ and then translated into values for $g$
and $M/\mu$.  The recursive nature is that each $\tilde\mu_m$ is computed
with only the Fock sectors above taken into account, by computing it in
a truncation where $\tilde\mu_m$ is the mass in the one-body sector and
the top sector has $\Nmax-m+2$ constituents.  In the top sector,
$\tilde\mu_{\Nmax}$ is just $\tilde M$.  For each sector in between,
the mass $\tilde\mu_m$ has been computed by solving the smaller problem where
$\tilde\mu_m$ was the mass in the one-body sector and the highest sector
had $\Nmax-m+1$ constituents with mass $\tilde M$.

To carry out this calculation, we define set of matrices $G^{(m)}$,
from $m=\Nmax$ down to 3, as
\be
G^{(m)}=\left[\tilde\mu_m^2 T^{(m)\prime}+V^{(m,m)\prime}-\tilde M^2 I^{(m)}
                -V^{(m,m+2)\prime}G^{(m+2)}V^{(m+2,m)\prime}\right]^{-1},
\ee
\be
G^{(\Nmax)}=\left[\tilde M^2 T^{(\Nmax)\prime}
                   +V^{(\Nmax,\Nmax)\prime}-\tilde M^2 I^{(\Nmax)}
                      \right]^{-1}.
\ee
The mass in the lowest sector is then simply
\be
\tilde\mu_1^2=\frac{1}{T^{(1)}}\left[\tilde M^2 -V^{(1,1)\prime}
    -V^{(1,3)\prime}G^{(3)}V^{(3,1)\prime}\right].
\ee
The coefficients for the wave-function expansions are constructed recursively 
from $m=3$ up to $\Nmax$ by
\be
\vec{c}^{\,(m)\prime}/c^{(1)}=G^{(m)}V^{(m,m-2)\prime}\vec c^{\,(m-2)\prime}/c^{(1)}.
\ee

\begin{figure}
\centering
\includegraphics[width=12cm]{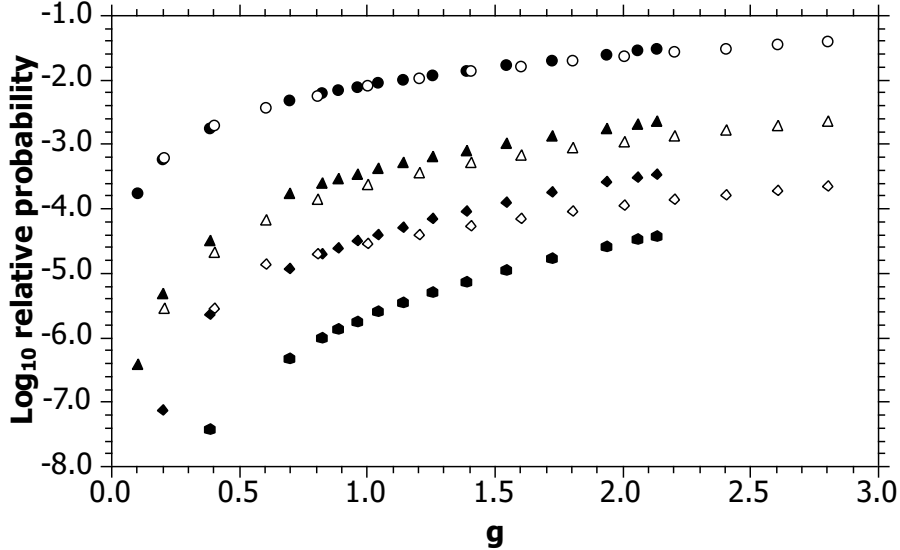}
\caption{Relative Fock-sector probabilities for the lowest mass eigenstate with
odd numbers of constituents, for both sector-dependent (closed symbols) and 
independent (open symbols), with the maximum number of constituents
$\Nmax=9$ and 7, respectively.  The sectors represented include three (circles),
five (triangles), seven (diamonds), and nine (hexagons) constituents.}
\label{fig:relprobsecdep}
\end{figure}

The relative probabilities are compared with those of the sector-independent
approach in Fig.~\ref{fig:relprobsecdep}.
The sector-dependent results do converge more slowly, with
respect to the Fock-space truncation, requiring $\Nmax=9$ to reach
the convergence found at $\Nmax=5$ for the sector-independent
calculation.  The slower convergence is to be expected, and
even desired, given that we expect the higher Fock states
to contribute more easily and to be more important as the
critical coupling is approached.  However, the relative
probabilities continue to show no critical behavior.

\section{Summary}  \label{sec:summary}

It is possible to understand the difference between ET and LF 
values of the critical coupling by taking the
different mass renormalizations into account.
However, calculation of the mass shift near the critical coupling
shows poor behavior of computed eigenstates.  The
relative probabilities of higher Fock states
do not show critical behavior; they should diverge
because the one-body probability should go to zero.
The use of a sector-dependent constituent mass
does not help.  The difficulty near the critical
coupling is avoided instead by extrapolation from
smaller coupling values.

The correct representation of the eigenstates near
critical coupling apparently requires a method without
Fock-space truncation.  In general, this would be
a coherent-state basis.  A specific implementation
might be the light-front coupled cluster method~\cite{LFCC}
with a nontrivial valence state, such as a linear
combination of the one-body and three-body states.

\acknowledgments
This work was done in collaboration with M. Burkardt and J.R. Hiller,
and was supported in part by the Minnesota Supercomputing Institute
of the University of Minnesota with allocations of computing
resources.

\end{document}